%% file: main.tex
\pdfoutput=1
\PassOptionsToPackage{dvipsnames}{xcolor}

\documentclass[sigconf, nonacm]{acmart}

\newcommand\vldbdoi{10.14778/3749646.3749677}

\newcommand\vldbavailabilityurl{}
\newcommand\vldbpagestyle{plain} 

\input{packages}

\begin{document}

\title{Semantic Integrity Constraints: Declarative Guardrails for AI-Augmented Data Processing Systems}

\author{Alexander W. Lee}
\affiliation{%
  \institution{Brown University}
}
\email{alexander\_w\_lee@brown.edu}

\author{Justin Chan}
\affiliation{%
  \institution{Brown University}
}
\email{juchan@brown.edu}

\author{Michael Fu}
\affiliation{%
  \institution{Brown University}
}
\email{michael\_fu@brown.edu}

\author{Nicolas Kim}
\affiliation{%
  \institution{Brown University}
}
\email{nicolas\_kim@brown.edu}

\author{Akshay Mehta}
\affiliation{%
  \institution{Brown University}
}
\email{akshay\_mehta@brown.edu}

\author{Deepti Raghavan}
\affiliation{%
  \institution{Brown University}
}
\email{deeptir@brown.edu}

\author{U\u{g}ur \c{C}etintemel}
\affiliation{%
  \institution{Brown University}
}
\email{ugur\_cetintemel@brown.edu}

\input{abstract}

\maketitle

\pagestyle{\vldbpagestyle}

\ifdefempty{\vldbavailabilityurl}{}{
\vspace{.3cm}
\begingroup\small\noindent\raggedright\textbf{PVLDB Artifact Availability:}\\
The source code, data, and/or other artifacts have been made available at \url{\vldbavailabilityurl}.
\endgroup
}

\input{introduction}
\input{architecture}
\input{specification}
\input{enforcement}
\input{observability}
\input{enterprise}
\balance%
\input{related}
\input{conclusion}
\input{acks}

\bibliographystyle{ACM-Reference-Format}
\bibliography{main}

\end{document}

%% file: packages.tex
\usepackage{cleveref}
\crefname{algoline}{line}{lines}
\Crefname{algoline}{Line}{Lines}
\crefname{algorithm}{Alg.}{Algs.}
\Crefname{algorithm}{Algorithm}{Algorithms}
\crefname{appendix}{App.}{App.}
\Crefname{appendix}{Appendix}{Appendices}
\crefname{corollary}{Corol.}{Corolls.}
\Crefname{corollary}{Corollary}{Corollaries}
\crefname{conjecture}{Conjecture}{Conjectures}
\Crefname{conjecture}{Conjecture}{Conjectures}
\crefformat{conjecture}{Conjecture~#2#1#3}
\crefmultiformat{conjecture}{Conjectures~#2#1#3}{\ and~#2#1#3}{, #2#1#3}{\ and~#2#1#3}
\crefname{definition}{Def.}{Defs.}
\Crefname{definition}{Definition}{Definition}
\crefname{figure}{Fig.}{Figs.}
\Crefname{figure}{Figure}{Figures}
\crefname{lemma}{Lemma}{Lemmas}
\Crefname{lemma}{Lemma}{Lemmas}
\crefname{lstlisting}{listing}{listings}
\Crefname{lstlisting}{Listing}{Listings} 
\crefname{proposition}{Prop.}{Props.}
\Crefname{proposition}{Proposition}{Propositions}
\Crefname{section}{Section}{Sections}
\crefname{section}{Sect.}{Sect.}
\crefname{subsection}{Sect.}{Sect.}
\Crefname{subsection}{Section}{Sections}
\crefname{subsubsection}{Sect.}{Sect.}
\Crefname{subsubsection}{Section}{Sections}
\crefname{table}{Table}{Tables}
\Crefname{table}{Table}{Tables}
\crefname{theorem}{Thm.}{Thms.}
\Crefname{theorem}{Theorem}{Theorems}

\usepackage{listings}


%% file: abstract.tex
\begin{abstract}
\emph{AI-augmented data processing systems (DPSs)} integrate large language models (LLMs) into query pipelines, allowing powerful semantic operations on structured and unstructured data. However, the reliability (a.k.a. trust) of these systems is fundamentally challenged by the potential for LLMs to produce errors, limiting their adoption in critical domains. To help address this reliability bottleneck, we introduce \emph{semantic integrity constraints (SICs)}---a declarative abstraction for specifying and enforcing correctness conditions over LLM outputs in semantic queries. SICs generalize traditional database integrity constraints to semantic settings, supporting common types of constraints, such as grounding, soundness, and exclusion, with both reactive and proactive enforcement strategies.

We argue that SICs provide a foundation for building reliable and auditable AI-augmented data systems. Specifically, we present a system design for integrating SICs into query planning and runtime execution and discuss its realization in AI-augmented DPSs. To guide and evaluate our vision, we outline several design goals---covering criteria around expressiveness, runtime semantics, integration, performance, and enterprise-scale applicability---and discuss how our framework addresses each, along with open research challenges.
\end{abstract}

%% file: introduction.tex
\section{Introduction}

Large language models (LLMs) have transformed the field of data management in the last couple of years, extending traditional systems with semantic processing capabilities over structured and unstructured data. In particular, several projects have integrated \emph{semantic operators} into database and data flow systems by augmenting traditional operators with LLMs~\cite{patel2024lotus,anderson2025aryn,liu2025palimpzest,lu2025vectraflow,shankar2024docetl,dai2024uqe,wang2025aop,urban2024caesura,russo2025abacus,aisql,madden2024databasesunbound}. We broadly refer to these systems as \emph{AI-augmented data processing systems (DPSs)}. While AI-augmented DPSs introduce novel ways to process data, their use of LLMs lead to potential reliability issues that manifest into phenomena like hallucinations. These reliability issues hinder the adoption of AI-augmented DPSs in high-stakes sectors, such as healthcare, law, and finance. Without guardrails, it will be difficult to adopt these systems fully.

LLM assertions have actively been explored for offline evaluations~\cite{trulens,zhang2023llmjudge,shankar2024spade,shankar2024evalgen,langchainai,llamaindex} and runtime guardrails~\cite{guardrailsai,rebedea2023nemo,dong2024buildingguardrails,singhvi2024dspyassertions,liu2024structuredoutput,loula2025syntactic}. These assertions invoke custom code, ML classifiers, or LLM judges to check an LLM's output with respect to various criteria. A related technique is constrained decoding, where constraints are applied during the LLM's generation to ensure that the output adheres to a specified structure~\cite{openaijson,dong2024xgrammar,willard2023outlines,zheng2024sglang,beurer2023lmql,guidance,koo2024automatabased,kuchnik2023relm}. While these approaches have proven their utility in enhancing LLM reliability, they are imperative and fragmented, limiting their usability.

To address these limitations, we propose \emph{semantic integrity constraints (SICs)} as a declarative, unified abstraction that extends traditional database integrity constraints to support AI-augmented DPSs. Unlike traditional integrity constraints that prevent invalid modifications to the database, SICs protect against erroneous outputs from semantic operators. Similar to their traditional counterparts, SICs make defining constraints for predominant use cases easy and more advanced cases possible, aligning with recent work that argues for specifications on LLM-based system components~\cite{stoica2024specifications}.

\begin{table*}[ht]
    \centering
    \caption{Success criteria and design goals for SICs}
    \footnotesize
    \begin{tabular}{p{3cm} p{13cm}}
        \toprule
        \textbf{Success Criterion} & \textbf{Description} \\
        \midrule
        Expressiveness & SICs support a range of constraint types---including domain, grounding, and soundness---through a  declarative interface (\Cref{sec:specification}) \\
        \midrule
        Executable Semantics & Each constraint has precise execution semantics—including what is checked, when it is enforced, and what actions are taken on failure—ensuring semantically consistent system behavior and predictable outcomes for users (\Cref{sec:enforcement}) \\
        \midrule
        System Integration & SICs integrate with query interfaces, optimizers, and execution engines in AI-augmented DPSs (\Cref{sec:architecture}) \\
        \midrule
        Performance Awareness & Enforcement strategy is selected based on cost and reliability across reactive and proactive methods (\Cref{sec:physopt}) \\
        \midrule
        Enterprise-Scale Applicability & SICs support observability, reuse, and auditability across workflows, with mechanisms for conflict detection and policy-driven enforcement at organizational scale (\Cref{sec:observability,sec:enterprise}) \\
        \bottomrule
    \end{tabular}
\label{tab:successcriteria}
\end{table*}

This paper presents our vision for SICs and sketches a general framework for integrating them into AI-augmented DPSs. To evaluate the feasibility and impact of this vision, we define a set of design goals and success criteria: (1) \emph{expressiveness}, to capture a broad range of common constraint types; (2) \emph{executable semantics}, with well-defined runtime behavior; (3) \emph{system integration}, via seamless support within data flow and query processing pipelines; (4) \emph{performance-aware enforcement}, balancing reliability and cost; and (5) \emph{enterprise-scale applicability}, including observability, constraint reuse, conflict detection, and auditability. These criteria, outlined in \Cref{tab:successcriteria}, were defined through our partnership with the Rhode Island Hospital. They guide our design and are reflected in the structure of the paper. \Cref{sec:architecture} outlines the proposed system architecture. \Cref{sec:specification} introduces the SIC interface and categorizes SICs across key enforcement classes aligned with real-world LLM use cases. \Cref{sec:enforcement} explores enforcement strategies, spanning logical and physical planning, and outlines cost-aware mechanisms. \Cref{sec:observability} discusses observability components. Finally, \Cref{sec:enterprise} presents open challenges in large-scale constraint management.

%% file: architecture.tex
\section{System Architecture}\label{sec:architecture}

We begin with an overview of the system architecture (\Cref{fig:architecture}) for supporting SICs and defer a more detailed discussion of the core components to later sections.  First, the user expresses their semantic query along with SICs via the \emph{query interface} (\Cref{sec:interface}). This interface forwards the query's text to the \emph{query parser}, which produces a corresponding logical query plan. Next, the \emph{logical optimizer} (\Cref{sec:logiopt}) applies logical rewrite rules to the initial plan, yielding an optimized logical plan. This plan is then passed to the \emph{physical optimizer} (\Cref{sec:physopt}), which considers different physical implementations of the plan and selects the optimal one. The resulting optimized physical plan is then executed by the \emph{execution engine}. To enable performance monitoring, an \emph{observability stack} is integrated with the execution engine (\Cref{sec:observability}). Finally, the \emph{constraint store} recommends and analyzes SICs (\Cref{sec:enterprise}).

%% file: specification.tex
\section{Constraint Specification}\label{sec:specification}

We suggest defining SICs declaratively, allowing users to focus on constraint logic while the optimizer determines the best execution method. We outline the main components of our proposed interface and detail classes of constraints relevant to semantic data processing. Additional surveys should be conducted to ensure usability.

\subsection{Interface}\label{sec:interface}

To provide a unified interface for queries and SICs, users declare constraints in the query language. We present a relational interface that extends pipe SQL syntax~\cite{shute2024pipesql} with semantic operators (e.g., map, aggregate, filter, join, top-$k$) and SICs; however, SICs are compatible with \emph{any} declarative AI-augmented DPS.

\textbf{Declaring Constraints.} SICs specify \emph{predicates} either on \emph{generated tuple attributes} from semantic operators or on the \emph{semantic operators} themselves. We make this distinction because while some operators (e.g., map, aggregate) generate attributes, other operators (e.g., filter, join, top-$k$) do not; yet, constraints should still be defined in either case. The syntax for declaring a constraint on an attribute involves a SQL expression starting with \texttt{ASSERT}, followed by a predicate involving the attribute's name, the type of constraint (\Cref{sec:classes}), and any necessary parameters. A constraint on an operator is similar. Users can also combine constraints via logical connectives (e.g., \texttt{ASSERT <pred1> AND <pred2>}).

\textbf{Retry Thresholds.} Users can define a retry threshold for each constraint. When a constraint associated with the operator (either on the operator or on an attribute generated by the operator) is violated, the DPS retries the operator until the output satisfies the constraint or the retry threshold is reached. To specify a retry threshold, users write \texttt{RETRY <threshold>} after the constraint.

\textbf{Failure Modes.} Failure modes (\texttt{CONTINUE}, \texttt{IGNORE}, or \texttt{ABORT}) are also associated with constraints and specify what the DPS should do when a failing output reaches the retry threshold. With \texttt{CONTINUE}, the DPS continues processing the query, allowing errors to propagate downstream. \texttt{IGNORE} specifies that the DPS should ignore invalid tuples in the processing (including dependent tuples, such as those that would be in the same aggregation window). Lastly, \texttt{ABORT} cancels the entire query. To specify a failure mode, the user writes \texttt{<mode> ON FAIL} after the retry threshold.

\textbf{End-To-End Example.} \Cref{lst:query} shows an example semantic query with various SICs. The query is declared over the \texttt{ehr\_table} (line~\ref{line:ehrtable}), a relation of electronic health records (EHRs). Each tuple in the relation corresponds to a patient, and we assume that it contains attributes for the patient's \texttt{ehr} and \texttt{dob}. The query first canonicalizes the \texttt{dob} attribute's value into \texttt{YYYY-MM-DD} (line~\ref{line:dob}) because the value could have been in another format. The expression involves a string prefixed with the letter ``p'', which defines a \emph{prompt string}. Prompt strings can specify natural language functions or predicates over attributes and are evaluated using LLMs. Line~\ref{line:age} computes the patient's age using standard SQL functions. Lines~\ref{line:physexam}-\ref{line:medhist} extract the patient's physical exam, lab results, and medical history from the \texttt{ehr}, and line~\ref{line:medhistsum} summarizes the patient's medical history. Lastly, the query filters tuples by selecting patients likely to have sepsis (line~\ref{line:sepsisfilter}). Here, the \texttt{AS} keyword is also used to alias the filter operator so that it can be referenced later in the query (line~\ref{line:sepsisfilterassert}).

The first constraint is declared on line~\ref{line:dobassert}, and the remaining constraints are specified on line~\ref{line:physexamlabresassert} and onward. If the summarization step for \texttt{med\_hist\_sum} (line~\ref{line:medhistsum}) fails the constraint on line \ref{line:medhistsumexclassert}, the DPS retries the operator once and then \texttt{CONTINUE}s to propagate the tuple even if it still fails the constraint again (line~\ref{line:retrycont}). Since no retry thresholds and failure modes are defined in the rest of the query, we assume that some default setting is used. We explain each constraint in more detail in the next section.

\begin{figure}[h]
\centering
\begin{lstlisting}[escapeinside={(*}{*)},caption={Semantic query with pipe SQL syntax and SICs.},label={lst:query}]
FROM ehr_table (*\label{line:ehrtable}*)
|> SET dob = p'canonicalize {dob} into YYYY-MM-DD' (*\label{line:dob}*)
|> ASSERT REGEXP_CONTAINS(dob, r'^\d{4}-(0[1-9]|1[0-2])-(0[1-9]|[12]\d|3[01])$') (*\label{line:dobassert}*)
|> EXTEND DATE_PART('year', AGE(CURRENT_DATE, dob::DATE)) AS age_yrs (*\label{line:age}*)
|> EXTEND EXTRACTIVE p'extract the patient''s admission physical exam from the {ehr}' AS phys_exam STRING (*\label{line:physexam}*)
|> EXTEND EXTRACTIVE p'extract the patient''s admission lab results from the {ehr}' AS lab_res STRING
|> EXTEND EXTRACTIVE p'extract the patient''s medical history from the {ehr}' AS med_hist STRING (*\label{line:medhist}*)
|> EXTEND ABSTRACTIVE p'summarize {med_hist}' AS med_hist_sum STRING (*\label{line:medhistsum}*)
|> WHERE p'the patient is likely to have sepsis based on their {age_yrs}, {phys_exam}, {lab_res}, and {med_hist}' AS sepsis_filter (*\label{line:sepsisfilter}*)
|> ASSERT phys_exam GROUNDED AND lab_res GROUNDED (*\label{line:physexamlabresassert}*)
|> ASSERT med_hist_sum GROUNDED AND LENGTH(med_hist_sum) < 1000 (*\label{line:medhistsumassert}*)
|> ASSERT med_hist_sum EXCLUDES p'test results' (*\label{line:medhistsumexclassert}*)
     RETRY 1 CONTINUE ON FAIL (*\label{line:retrycont}*)
|> ASSERT sepsis_filter SOUND (*\label{line:sepsisfilterassert}*)
\end{lstlisting}
\Description{Semantic query with pipe SQL syntax and SICs.}
\end{figure}

\begin{figure*}[t]
  \centering
  \includegraphics[trim=0 14cm 0 0, clip, width=\textwidth]{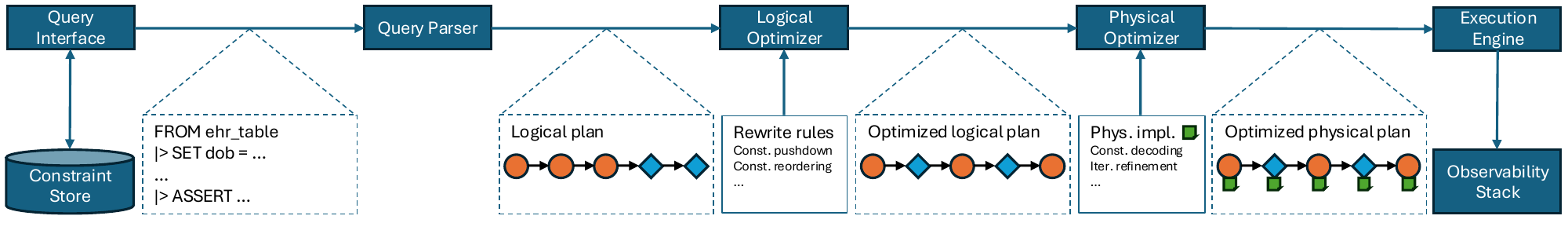}
  \caption{System architecture for SICs. Orange circles and turquoise diamonds represent semantic operators and SICs, respectively. Green squares indicate annotations that describe the physical implementation of each node.}
  \Description{System architecture for SICs. Orange circles and turquoise diamonds represent semantic operators and SICs, respectively. Green squares indicate annotations that describe the physical implementation of each node.}
  \label{fig:architecture}
\end{figure*}

\subsection{Constraint Classes}\label{sec:classes}

We categorize SICs into different classes (\Cref{tab:classes}) that are designed to capture the predominant use cases for LLM constraints. We motivate each constraint class using concrete examples from the EHR query (among others) and briefly outline their implementation.

\textbf{Domain Constraints.} In many queries, users want generated attributes to belong to a specified \emph{domain}. This includes use cases such as enforcing that a value conforms to a certain data type or structure, belongs to a set of values, or is less than a certain length. For example, one may check whether the predicted prescription dose from a semantic map is within a range of possible values. We extend traditional domain constraints to semantic data processing. In \Cref{lst:query}, line~\ref{line:dobassert} and the second conjunct on line~\ref{line:medhistsumassert} specify domain constraints, restricting the \texttt{dob} to the regex for \texttt{YYYY-MM-DD} and the summary's length to less than 1000 characters, respectively. Type declarations (e.g., \texttt{STRING}) of generated attributes implicitly define domain constraints. This class of SICs comprises the majority of constraint use cases~\cite{liu2024structuredoutput} and is enforced deterministically.

\textbf{Inclusion and Exclusion (IE) Constraints.} Users sometimes expect that an attribute generated by an operator \emph{includes} or \emph{excludes} certain terms, content, topics, or abstract features such as tones or styles. For example, given a semantic map that summarizes an EHR, users may want to exclude all the patient's PII from the summary while including the names of attending physicians. IE constraints address this class of predicates. Line~\ref{line:medhistsumexclassert} defines an exclusion constraint on \texttt{med\_hist\_sum}, which excludes test results from the medical history summary. When an IE constraint is defined with a prompt string (as in the example), LLMs are used to check for a semantic match. Users can also specify IE constraints using strings, regexes, or even a collection of them (e.g., a set of toxic words to exclude from the output). In such cases, deterministic methods are used to evaluate matches.

\textbf{Grounding Constraints.} Semantic operators, such as maps or aggregates, can extract data or summarize from their inputs. However, since LLMs can hallucinate, users want to ensure that the generated data is \emph{grounded} in source data. Grounding constraints enforce such invariants. For example, line~\ref{line:physexamlabresassert} ensures that the extracted \texttt{phys\_exam} and \texttt{lab\_res} are \emph{exactly contained} in the source EHR, while the first conjunct on line~\ref{line:medhistsumassert} asserts that the summarized \texttt{med\_hist\_sum} is \emph{factually consistent} with the source EHR. In particular, the last grounding constraint implicitly defines a grounding constraint on \texttt{med\_hist}, since \texttt{med\_hist\_sum} is derived from \texttt{med\_hist}, which, in turn, is derived from \texttt{ehr}. Formally, if a grounding constraint is defined on an attribute, the DPS will recursively enforce grounding constraints on all ancestor attributes in its lineage. Thus, each constraint only needs to compare its operator's input and output to ensure grounding in the source data. A grounding constraint is implemented in one of two ways, depending on whether the specified attribute is generated via an \emph{extractive} (map) or \emph{abstractive} (map or aggregate) operator. In the extractive case, the constraint is implemented deterministically and ensures that the operator's output is precisely contained in its input. In the abstractive case, the constraint is implemented stochastically (e.g., with an LLM or a fact-checking model~\cite{tang2024minicheck}) and ensures that the operator's summarized output is factually consistent with the input. Users can annotate relevant operators with the \texttt{EXTRACTIVE} (lines~\ref{line:physexam}-\ref{line:medhist}) or \texttt{ABSTRACTIVE} (line~\ref{line:medhistsum}) keywords to inform the DPS of their semantics; otherwise, the DPS infers these annotations by reasoning about the operator's prompt string with an LLM.

\begin{table*}[t]
    \centering
    \caption{Summary of SIC classes. The constraint classes are derived from the taxonomy introduced in prior work~\cite{liu2024structuredoutput} and the use cases identified from our collaboration with the Rhode Island Hospital.}
    \footnotesize
    \begin{tabular}{p{2cm} p{5cm} p{9cm}}
        \toprule
        \textbf{Class} & \textbf{Description} & \textbf{Examples}\\
        \midrule
        Domain & Value is from a specified domain & Data types, structured output, multiple choices, numerical ranges, or length constraints\\
        \midrule
        Inclusion/Exclusion & Value incl/excl terms, content, tones, styles, etc. & Inclusion of attending physician names or exclusion of PII from EHR summaries\\
        \midrule
        Grounding & Value is grounded in the source values & Extracted test results or summaries are factual w.r.t. source EHRs\\
        \midrule
        Soundness & LLM's reasoning process is logically sound & Predictions of patient diagnoses are based on sound reasoning\\
        \midrule
        Relevance & Value is relevant to the given task & Extracted histories from EHRs are medical (desired) rather than social (undesired)\\
        \midrule
        Assertions & Arbitrary predicate holds in the data flow & All included URLs in generated research reports lead to valid websites\\
        \bottomrule
    \end{tabular}\label{tab:classes}
\end{table*}

\textbf{Soundness Constraints.} To check whether the reasoning of an operator is \emph{logically sound}, users can declare soundness constraints. Although applicable to all operators, this constraint class is particularly useful for operators that do \emph{not} generate attributes, such as filter, join, or top-$k$. To enforce soundness constraints, the DPS uses LLMs to check the operator's reasoning, e.g., chain-of-thought (CoT)~\cite{wei2022cot}. As an example, line~\ref{line:sepsisfilterassert} declares a soundness constraint on the sepsis filter. Assume that the query earlier retrieved relevant sepsis-related medical information (omitted for brevity). This retrieved information is considered to describe the symptoms of sepsis and the diagnostic rationale. Under the hood, the DPS prompts the filter's LLM to return its CoT in addition to a boolean value that represents the selection result. The CoT first lists the premises, i.e., the patient's medical condition and the retrieved medical information relevant to sepsis. The premises are followed by the reasoning steps that lead to the selection result. Then, the input and output of the filter's LLM are given to an LLM judge (or several of them) to ensure that the premises are grounded in the input and that the reasoning steps are valid. The DPS can also use more sophisticated CoT prompting and verification techniques, which leverage structured natural language formats~\cite{ling2023deductivecot}.

\textbf{Relevance Constraints.} These constraints validate if an attribute's value is \emph{relevant} for the given operator's task; i.e., they verify whether the underlying LLM followed the instructions prompted by the user. Since relevance constraints are always appropriate, the DPS can enforce them on all generated attributes \emph{by default}. We assume that this is the case in the example query. For instance, the DPS checks whether the medical history extracted in line \ref{line:medhist} is in fact the patient's medical history rather than something irrelevant, such as their social history. These constraints can be evaluated by passing the operator's prompt, input, and output to an LLM judge.

\textbf{Assertions.} Like assertions in traditional integrity constraints, this class of constraints enables users to specify \emph{arbitrary predicates} that must hold in the data flow. The DPS cannot offer many optimizations for assertions due to their black-box nature; however, such constraints give users the flexibility to express conditions that are not supported by the classes described previously.

%% file: enforcement.tex
\section{Constraint Enforcement}\label{sec:enforcement}

Constraint enforcement consists of two phases: \emph{detection} and \emph{recovery}. First, a constraint \emph{detects} any violations in the operator's output. If a violation is detected, the operator attempts to \emph{recover}. This distinction helps us reason about optimizing constraint enforcement, as discussed in this section.

\subsection{Logical Optimization}\label{sec:logiopt}

The logical optimizer takes as input a logical query plan represented as a data flow graph, where each node is either an operator or a constraint. The goal of the logical optimizer is to minimize the cost of the plan by applying \emph{rewrite rules} to semantic operators and constraints. Here, we refer to cost as any arbitrary cost metric, such as latency or monetary cost. To keep the presentation focused, we only discuss logical optimizations for constraints.

\textbf{Constraint Pushdown.} First, the optimizer moves all constraints to their earliest evaluation point in the data flow (i.e., immediately following their corresponding operators). This allows the DPS to detect and recover from constraint violations as soon as possible rather than unnecessarily processing erroneous outputs. The optimizer also appends the textual representation of each constraint to their corresponding operator's LLM prompt, ensuring that the underlying model is aware of the constraints on its results.

\textbf{Constraint Reordering.} The optimizer then reorders the SICs for the same operator. Following prior work~\cite{abadi2003aurora}, the reordering is based on the cost and selectivity of each constraint, where selectivity is defined as the probability that a tuple satisfies the constraint's predicate. Cost and selectivity estimates can be computed from sampled data. We discuss more details on estimation in \Cref{sec:planselection}.

\subsection{Physical Optimization}\label{sec:physopt}

Given an optimized logical plan, the physical optimizer considers different implementations of the plan and returns the optimal one. We first discuss approaches to constraint enforcement and then outline ideas for optimized plan selection.

\subsubsection{\textbf{Enforcement Implementations}}\label{sec:implementations}

Constraints are enforced \emph{reactively} or \emph{proactively}, based on whether enforcement occurs \emph{after} or \emph{during} response generation. While all constraints can be enforced reactively, only some can be enforced proactively.

\textbf{Reactive Enforcement.} After the LLM produces its entire response, there are different methods to detect a constraint violation. For example, a grounding constraint on an extractive operator can simply check if the operator's output attribute is contained in the input. On the other hand, a grounding constraint on an abstractive operator can invoke a single LLM judge~\cite{zhang2023llmjudge}, a fact-checking model~\cite{tang2024minicheck}, a compound AI system~\cite{chern2023factool}, or anything in between. Similarly, there are different approaches for an operator to recover from a violation by retrying. The common approach used in current systems is \emph{iterative refinement with feedback}~\cite{singhvi2024dspyassertions,shankar2024docetl,madaan2023selfrefine}. Furthermore, the operator can retrieve additional semantically similar few-shot examples~\cite{brown2020fewshot} of prior input-output pairs that satisfied the constraint. These examples can be stored in a cache built on a vector index. Other approaches include retrying with more powerful models, larger ensemble sizes, or higher thinking budgets~\cite{anthropicthinking,geminithinking}.

\textbf{Proactive Enforcement.} Proactively enforcing constraints can lower costs significantly. For example, \emph{constrained decoding}~\cite{openaijson,dong2024xgrammar,willard2023outlines,zheng2024sglang,beurer2023lmql,guidance,koo2024automatabased,kuchnik2023relm} pushes enforcement into the LLM's decoding process, masking invalid tokens to ensure compliance by construction. Various libraries~\cite{openaijson,dong2024xgrammar,willard2023outlines,zheng2024sglang,guidance} support JSON Schema~\cite{jsonschema} adherence in LLM outputs by converting the schema into a grammar and then an automaton for token masking. However, constrained decoding is not limited to grammar-based approaches. For instance, using a suffix automaton is more efficient than regex to constrain an LLM to output a substring of its input~\cite{koo2024automatabased}, enabling proactive enforcement of grounding constraints for extractive operators. Yet, we are not aware of any existing library that offers this functionality.

An open challenge is to identify which other SICs can be enforced via constrained decoding. Moreover, developing a framework to apply efficiently various constraints on different output attributes also remains unresolved. Prior research~\cite{beurer2023lmql} has seen some advancement but remains slower than modern grammar-based techniques~\cite{zheng2024sglang}. Nonetheless, it provides a foundation for future investigation. Importantly, we suggest that the DPS should limit constrained decoding to simple constraints (seen by the LLM during training) to avoid sampling unlikely token distributions. This rationale also underlies why constraints are included in the LLM's prompt before decoding.

Besides deterministic constrained decoding, the DPS can proactively enforce some stochastic constraints on partial outputs. For instance, the DPS can enforce a grounding constraint on an abstractive operator by invoking a fact-checking model~\cite{tang2024minicheck} after each sentence is generated. While recent research~\cite{loula2025syntactic} explores these ideas, they do not focus on efficient optimizations, such as \emph{asynchronously} detecting violations in partial streamed outputs. This optimization is similar to previous work~\cite{santhanam2024alto}, except that the operator now needs to recover if its partial output violates a constraint. Upon a violation, the operator can backtrack to the decoding step right before the violation; though, alternative techniques like re-prompting with partial feedback are also worth considering.

\subsubsection{\textbf{Optimized Plan Selection}}\label{sec:planselection}

For a \emph{fixed} set of constraints and their implementations, the physical optimizer can adapt existing techniques~\cite{liu2025palimpzest,russo2025abacus} to select an optimal query plan defined by user-specified cost or reliability (a.k.a. quality) thresholds (omitted from the query interface for brevity). Specifically, rather than measuring the \emph{reliability of an operator's implementation} in terms of an oracle, reliability is instead defined by adherence to all of the operator's constraints (after necessary retries). The challenge is selecting each constraint's implementation in the first place.

As illustrated in~\Cref{sec:implementations}, each constraint's implementation has unique cost-reliability trade-offs, which allow users to specify additional thresholds to control them. While existing \emph{cost} estimation approaches can be extended to account for each operator's constraints, estimating the \emph{reliability} of the constraints is a more difficult task. Formally, the \emph{reliability of a constraint's implementation} is defined as the \emph{precision} and \emph{recall} of detecting a violation. A deterministic constraint is completely reliable (i.e., 100\% precision and recall) and requires no estimation, while a stochastic constraint is \emph{not} completely reliable due to invoking models. Estimating the reliability of stochastic constraint implementations is challenging. First, such estimates require ground-truth labels from human annotators. Furthermore, reliability is a function of both the implementation and its input, and this input, in turn, depends on the upstream query plan and the underlying data. We present some initial ideas for constraint reliability estimation.

One strategy is to use \emph{historical} data and workloads to estimate the precision and recall of different implementations for each constraint, providing these estimates to the physical optimizer for use at runtime. However, this approach neglects the dependencies mentioned earlier, leading to possibly noisy estimates. Since metrics are aggregated across diverse query contexts, they do not effectively reflect interactions with upstream operators. Furthermore, data and workload drift may make these estimates obsolete. It is also costly for individual organizations to annotate execution traces.

An alternative approach is to develop a \emph{shared} foundation model for organizations, where given representations of (1) a constraint's implementation, (2) the implementation of the query plan up until the constraint, and (3) the underlying data, the model predicts the precision and recall for the constraint's implementation. Organizations can optionally fine-tune the model for their use cases. Recent work on foundation database models~\cite{wehrstein2025foundationdbmodels} are applicable here, with extensions to address the complexities of semantic processing.

A third approach is to use the model's \emph{confidence score} as a proxy for constraint reliability, allowing the optimizer to estimate the reliability (i.e., confidence) of a constraint's implementation at runtime based on sampled data. Rather than defining precision and recall thresholds, users specify confidence thresholds instead. While many conventional classifiers inherently provide confidence scores, they can also be derived from LLMs by applying softmax to the ``True'' and ``False'' token logits. The difficulty lies in achieving calibrated confidence scores that accurately represent the implementation's true correctness probability. Studies~\cite{kadavath2022lmsmostlyknow,liu2025calibratingguardrails} indicate that methods like temperature scaling~\cite{guo2017tempscaling} can enhance calibration in binary classification tasks. Though, further research is needed to obtain highly-calibrated constraints across varying domains.

%% file: observability.tex
\section{Observability}\label{sec:observability}

After a user submits a query to the DPS, the observability stack enables the user to understand the performance of their query's operators and constraints. The stack has three components. The \emph{observability interface} allows the user to view relevant metrics (e.g., cost, reliability) and inspect query results. The \emph{observability store} maintains metrics and outputs from each stage in the query. Lastly, the \emph{labeling interface} enables human annotators to provide ground-truth labels to compute the precision and recall of each stochastic constraint. While methods such as LLM-as-a-judge offer a scalable approximation of human preferences~\cite{zhang2023llmjudge} and have become widespread~\cite{shankar2024spade,shankar2024evalgen,shankar2024docetl,singhvi2024dspyassertions,madaan2023selfrefine,guardrailsai,rebedea2023nemo,langchainai,dong2024buildingguardrails,llamaindex,trulens}, they are not entirely reliable. As a result, labeling interfaces remain essential. We detail each component of the stack below.

\textbf{Observability Interface.} Given a user-specified query ID, the observability interface immediately shows metrics such as cost, operator reliability, and constraint selectivity. However, to observe the reliability of stochastic constraints, the user must submit a labeling request for their query. Once the labeling is complete, the interface displays the precision and recall of each constraint. The interface can also indicate whether a constraint is deterministic or stochastic; though, the query interface can expose this distinction through the \texttt{EXPLAIN} command as well. Moreover, the observability interface allows the user to inspect individual output tuples. A tuple is flagged if any upstream operator violated a constraint but \texttt{CONTINUE}ed to propagate the error. The user can also view the tuple's lineage for more fine-grained debugging.

\textbf{Observability Store.} The observability store is modeled as additional relations in the DPS. It maintains the lineage of each query, enabling systematic debugging and performance monitoring. To obtain human annotations, the DPS can draw concepts from crowd-sourced databases~\cite{franklin2011crowddb,marcus2011crowdsourceddbs,park2012deco}, such as crowd-sourced attributes and user interface generation; however, the DPS need not implement these ideas in their full generality. For example, suppose the DPS stores a relation where each tuple represents an invocation of a constraint. The relation includes attributes for the constraint's input and output (i.e., predicted label) among others (e.g., query ID). In particular, the relation also has a crowd-sourced attribute for the true label (initially null).  When the user initiates a labeling request, the observability interface submits a query to the DPS, which computes (over a sample) the precision and recall of each constraint for the given query ID. The query populates each sampled true label by sending tasks to a crowd-sourcing platform.

\textbf{Labeling Interface.} A key aspect of crowd-sourced databases is their ability to create user interfaces automatically for crowd workers based on declarative queries. Here, the DPS produces a labeling interface for annotators, which displays a description of the constraint and its inputs. The annotator's role is to determine if the inputs satisfy the constraint.

%% file: enterprise.tex
\section{Enterprise-Wide Constraints}\label{sec:enterprise}

The introduction of SICs into an enterprise environment presents both new opportunities and challenges. To effectively manage SICs within an organization, we propose the development of a \emph{constraint store} and outline its key desired features.

\textbf{Recommendations.} Although declaring SICs is relatively simple, determining the applicable constraints is more difficult. This requires anticipating the potential failure cases for each semantic operator in the query. Recent systems have been developed to assist users in formulating constraints~\cite{shankar2024spade,shankar2024evalgen}, but these approaches typically generate suggestions based solely on the contents of the user's query. In enterprise settings, constraint recommendations can be greatly enriched. Organizations typically enforce domain-specific rules that necessitate certain constraints for all pertinent queries. For instance, a hospital might require excluding PII in queries unless essential for analysis. Moreover, different teams within the same organization might use similar queries with shared constraints. Thus, the constraint store needs to capture and manage common constraints automatically organization-wide, facilitating their reuse and recommendation in related queries. We suggest using text embeddings~\cite{openaiembeddings,reimers2019sentencebert} and graph embeddings~\cite{xu2018gin,kipf2017gcn,velickovic2018gat} to represent constraints and queries. This approach enables similarity searches to identify relevant constraints, which can then be refined by a reranker and LLM to match the user's context.

\textbf{Conflicts.} Conflicts can arise when queries contain multiple SICs, such as from accumulating shared constraints across an organization's workloads. For example, a query that requests a grounded list of the patient's medical conditions with full details while also requiring the exclusion of PII may encounter conflicts when the relevant clinical information (e.g., diagnosis dates, provider names, hospital locations) are themselves considered PII, making it infeasible to satisfy the constraints without redaction or abstraction. Likewise, domain constraints over disjoint value sets can lead to inevitable failures. The constraint store allows for conflict analysis using static reasoning and LLM-based methods, issuing warnings. Users can resolve conflicts by adjusting priorities (e.g., making one constraint as soft) or relaxing scope as needed. This approach builds on early research in active databases~\cite{widom1996starburst,hanson1996design,dayal1988hipac}, which suggested techniques for detecting rule conflicts and execution issues, such as non-termination, before execution~\cite{aiken1992behavior}.

%% file: related.tex
\section{Related Work}

Database integrity constraints were first described in \citeauthor{codd1970relational}'s relational model~\cite{codd1970relational} and later implemented in INGRES~\cite{held1975ingres,stonebraker1975implementationofic} and System R~\cite{astrahan1976systemr,chamberlin1976sequel2,eswaran1975functional}. Since then, constraints have been supported by all major databases. ML assertions were introduced to monitor and enhance model performance by applying arbitrary functions to model outputs to find potential errors~\cite{kang2020modelassertions}. Recent developments broaden assertions to LLMs by employing LLM judges and ML classifiers for offline evaluations~\cite{trulens,zhang2023llmjudge,shankar2024spade,shankar2024evalgen,langchainai,llamaindex} and runtime guardrails~\cite{guardrailsai,rebedea2023nemo,dong2024buildingguardrails,singhvi2024dspyassertions,liu2024structuredoutput}. Tools like \textsc{spade}~\cite{shankar2024spade} and \textsc{EvalGen}~\cite{shankar2024evalgen} guide users by suggesting relevant assertions and their implementations (e.g., Python functions, LLM grader prompts). These tools complement SICs and can instead suggest declarative constraints to boost enforcement reliability and efficiency. Our vision for enterprise constraints also draws inspiration from these efforts. DSPy Assertions enable users to implement assertions in LLM pipelines as functions, which compile optimized prompts and iteratively refine outputs at runtime~\cite{singhvi2024dspyassertions}. While SICs use iterative refinement, they do not currently support prompt optimization capabilities, although they can be added. A related area of work is controlled generation, which encompasses constrained decoding~\cite{openaijson,dong2024xgrammar,willard2023outlines,zheng2024sglang,beurer2023lmql,guidance,koo2024automatabased,kuchnik2023relm} and model-based constraints~\cite{loula2025syntactic}. To the best of our knowledge, there is no unified framework that offers declarative and optimized constraint enforcement.

Several recent systems have augmented traditional database operators with LLMs~\cite{patel2024lotus,anderson2025aryn,liu2025palimpzest,lu2025vectraflow,shankar2024docetl,dai2024uqe,wang2025aop,urban2024caesura,russo2025abacus,aisql,madden2024databasesunbound}. So far, DocETL is the only AI-augmented DPS to provide validation as a first class citizen~\cite{shankar2024docetl}. They enable users to express simple Python statements and validation prompts for iteratively refining operator outputs. However, their approach to validation is largely imperative. LOTUS~\cite{patel2024lotus} and Palimpzest~\cite{liu2025palimpzest,russo2025abacus} do not support constraints, but they allow users to specify quality thresholds relative to oracles.

%% file: conclusion.tex
\section{Conclusions}

SICs offer a unified, declarative foundation for improving the reliability of AI-augmented data processing. By generalizing traditional integrity constraints to the semantic behaviors of LLM-augmented operators, SICs bridge the gap between declarative query processing and the uncertainty of generative models. Their integration into the planning and execution layers of modern DPSs enables not only reliable enforcement but also cost-aware optimization, system-wide coordination, and organizational reuse. Through this framework, we outline a principled path toward building trustworthy AI-augmented systems capable of supporting mission-critical workflows. SICs are not only a guardrail mechanism, but also a catalyst for a new class of optimizations, abstractions, and research directions at the intersection of databases and LLMs.

%% file: acks.tex
\begin{acks}
This material is based upon work supported by the National Science Foundation Graduate Research Fellowship Program under Grant Nos 2439559 and 2040433. Any opinions, findings, and conclusions or recommendations expressed in this material are those of the authors and do not necessarily reflect the views of the National Science Foundation. We thank Shu Chen, Duo Lu, Malte Schwarzkopf, Weili Shi, and the VectraFlow team for their valuable feedback.
\end{acks}